\documentclass[prl,twocolumn,nofootinbib,showpacs]{revtex4}
\usepackage{graphicx}
\usepackage{epsfig}
\usepackage{dcolumn}
\usepackage{bm}
\usepackage{amsmath}
\usepackage{amsbsy}
\usepackage{amssymb}
\usepackage{lscape}
\usepackage[usenames]{color}

\newcommand{\bra}[1]{\ensuremath{\langle{#1}|\,}}
\newcommand{\ket}[1]{\ensuremath{\,|{#1}\rangle}}

\begin{document}


\title{Theory of the optical absorption of light carrying orbital angular momentum by semiconductors}

\author{G.\ F.\ Quinteiro and P.\ I.\ Tamborenea}
\affiliation{
\mbox{Departamento de F\'{\i}sica ``J.\ J.\ Giambiagi'', Universidad de
Buenos Aires}
\mbox{Ciudad Universitaria, Pabell\'on I, C1428EGA Ciudad de Buenos Aires,
Argentina}}

\date{\today}

\begin{abstract}
We develop a free-carrier theory of the optical absorption of light carrying 
orbital angular momentum (twisted light) by bulk semiconductors.
We obtain the optical transition matrix elements for Bessel-mode
twisted light and use them to calculate the wave function of photo-excited
electrons to first-order in the vector potential of the laser.
The associated net electric currents of first and second-order on the field
are obtained. 
It is shown that the magnetic field produced at the center of the beam 
for the $\ell=1$ mode is of the order of a millitesla, and could therefore 
be detected experimentally using, for example, the technique of 
time-resolved Faraday rotation.

\end{abstract}

\maketitle


It is well known from classical electromagnetism that light can
carry spin and orbital angular momentum. 
While the former was
detected for the first time in the 1930s~\cite{bet}, the latter
became widely available for experimental study only recently after
the work of Allen {\it et al.}~\cite{all-bei-spr} 
In a seminal
paper, those authors showed that light carrying an integer amount of
\textit{orbital} angular momentum ($\hbar\,l$, with $l$ an integer)
may be generated in the laboratory using conventional laser beams.
Since then, research on the subject of light carrying orbital
angular momentum (OAM), or twisted light
(TL)~\cite{mol-tor-tor,pad-cou-all,all-pad-bab} has spanned a large
number of areas, namely, generation of beams~\cite{pad-cou-all},
interaction with mesoscopic particles (optical
tweezers)~\cite{bar-tab, all, fri-nie-hek-rub}, entanglement with
spins for potential applications in quantum information
processing~\cite{mut-str}, interaction with atoms and
molecules~\cite{dav-and-bab,ara-ver-cla}, cavity-QED~\cite{ala-bab},
and interaction with Bose-Einstein
condensates~\cite{and-ryu-cla,sim-nyg-hu}. 
Nevertheless, the
interaction with solid state systems, although potentially important
for technological applications, has not been explored so far. 
In
this Letter, we present the first theoretical predictions about the
interaction of TL with bulk semiconductors. 
We consider band-to-band
transitions, i.e.\ optical transitions with light frequencies above
the bandgap, so that free carriers rather than excitons are
produced. 
We show that there is a transfer of OAM between the light
and the photoexcited electrons so that a net electric current
initially confined to the beam area appears. 
The magnetic field
induced by these photocurrents is estimated.

A beam of TL presents an azimuthal phase dependence --- helical wavefront ---
responsible for the OAM,
and a radial dependence of the \textit{Laguerre-Gaussian}(LG) or
\textit{Bessel mode} type.
We will focus on the Bessel modes, but our results are applicable to the
LG-mode with slight changes.
The vector potential in the Coulomb gauge with cylindrical coordinates
$\{r_\parallel, \phi , z\}$ is~\cite{jau}
\begin{eqnarray}
\label{Eq_A}
    \mathbb{A}_{{\bf q}\,l\,\pm}(\textbf{r},t)
&=& A_0 \, e^{i(q_z z- \omega t)}
    \, \left[ \boldsymbol{\epsilon}_{\pm}
    J_{l}(q_\parallel r_\parallel) e^{i l \phi} \mp  \right. \nonumber \\
&&  ~~~ \left. i \,\hat{z}
    \frac{q_\parallel}{q_z} J_{l\pm 1}(q_\parallel r_\parallel)
    e^{i (l \pm 1) \phi} \right] + c.c. \,,
\end{eqnarray}
with polarization vectors $\boldsymbol{\epsilon}_{\pm} = \hat{x}\pm
i\hat{y}$, Bessel functions $J_l$, and parameters $q_\parallel \ll
q_z$.

Semiconductors are solids which at zero temperature have the highest
occupied (valence) and lowest empty (conduction) energy bands
separated by a gap $E_g$. In this respect, they are closer to
insulators than to metals. However, in typical semiconductors $E_g
\simeq 1 \, \mbox{eV}$, making possible the transitions between the
valence and conduction bands by optical excitation. In a crystalline
semiconductor, electrons in the valence and conduction bands,
denoted by $\lambda=\{v,c\}$, occupy the Bloch states
\mbox{$\varphi_{\lambda \textbf{k}}(\textbf{r}) = L^{-3/2}
e^{i\,\textbf{k} \,\cdot \,\textbf{r}} u_{\lambda
\textbf{k}}(\textbf{r})$}, with $u_{\lambda \textbf{k}}(\textbf{r})$
a cell-periodic function (lattice constant $a$),
$\hbar\,\textbf{k}$ the crystal momentum, and $L$ the linear size of
the semiconductor. This simplified two-band model grasps the main
features of a semiconductor\cite{hau-koc}, and can be considered a
good model of a real bulk system under the following two conditions:
$i)$ an applied strain splits the heavy-hole and light-hole valence
bands; $ii)$ only circularly polarized light is considered,
producing optical transitions of only one electron spin type between
valence and conduction bands.

We now proceed to develop an analytical description of the interband
transitions in a two-band model of a bulk semiconductor induced by
twisted light. Our analysis tackles the coherent optical excitation
to conduction-band states, i.e.\ we consider the case
$\hbar\omega>E_g$. In this regime the creation of excitons is
negligible and one only has to take into account the free carriers
transferred by optical excitation from the valence to the conduction
band~\cite{hau-koc}. The lowest-order contribution to the
light-matter interaction, using the minimal-coupling Hamiltonian, is
$ H_{I} = - (Q/m)\, \textbf{p} \cdot \textbf{A}_{{\bf
q}\,l\,\sigma}(\textbf{r})$ with $\textbf{A}_{{\bf
q}\,l\,\sigma}(\textbf{r})$ the transverse or $xy$ part of
$\mathbb{A}_{{\bf q}\,l\,\pm}(\textbf{r},t)$ [Eq.\ (\ref{Eq_A})],
and $\{\textbf{p}, m, Q\}$ the momentum operator, mass, and charge
of the particle involved. The complete semiclassical Hamiltonian ---
operators for electrons and classical variables for the light field
--- consists then of two terms, the bare electron energy and the
interaction $H_I$
\begin{eqnarray*}
    H &=& \sum_{\lambda {\textbf{k}}} E_{\lambda \textbf{k}}
        a^\dagger_{\lambda \textbf{k}}
        a_{\lambda \textbf{k}} +\nonumber \\
    &&  \sum_{\lambda^\prime {\textbf{k}}^\prime\lambda {\textbf{k}}}
        \frac{-Q}{m} \bra{\lambda^\prime \textbf{k}^\prime}\textbf{p} \cdot
        \textbf{A}_{{\bf q}\,l\,\sigma}(\textbf{r})\ket{\lambda \textbf{k}}
     \, a^\dagger_{\lambda^\prime \textbf{k}^\prime}
        a_{\lambda \,\textbf{k}} \,,
\end{eqnarray*}
with $a^\dagger_{\lambda^\prime \textbf{k}^\prime}/a_{\lambda
\textbf{k}}$ the creation/annihilation operators for electrons in a
Bloch state of band $\lambda=v,c$ and quasi-momentum
$\hbar\,\textbf{k}$, and $E_{\lambda \textbf{k}}$ its bare energy.
The action of $\textbf{p}=-i\,\hbar\, \boldsymbol{\nabla}$ onto the
wave function \mbox{$\varphi_{\lambda \textbf{k}}(\textbf{r}) = \langle
\textbf{r}\ket{\lambda \textbf{k}}$} yields two terms: $\textbf{p}
\,\varphi_{\lambda \textbf{k}}(\textbf{r}) = \hbar\, \textbf{k}
\,\varphi_{\lambda \textbf{k}}(\textbf{r}) -i\,\hbar\,L^{-3/2}
e^{i\textbf{k} \,\cdot\, \textbf{r}} \nabla u_{\lambda
\textbf{k}}(\textbf{r})$. For interband transitions -- our interest
-- the first term is negligible; then
\begin{eqnarray*}
    H_{I}|_{\lambda^\prime \textbf{k}^\prime\lambda \textbf{k}}
        &=& \frac{i\hbar}{L^{3}} \frac{Q}{m} \int_{L^3} d\textbf{r} \,
            e^{-i (\textbf{k}^\prime-\textbf{k})\cdot \textbf{r}} \\
        &&  u^*_{\lambda^\prime \textbf{k}^\prime}(\textbf{r})
            [\textbf{A}_{{\bf q}\,l\,\sigma}(\textbf{r}) \cdot
            \nabla] u_{\lambda \textbf{k}}(\textbf{r}).
\end{eqnarray*}
This expression can be simplified by using a standard procedure in
solid state physics. The integration over the complete system is
replaced by an integration over a unit cell and a sum over all cells
\begin{eqnarray*}
    H_{I}|_{\lambda^\prime \textbf{k}^\prime\lambda \textbf{k}}
        &=& -\frac{Q}{m}
            \frac{1}{N}\,\sum_c\,
            e^{-i (\textbf{k}^\prime-\textbf{k})\cdot \textbf{R}_c}
            \textbf{A}_{{\bf q}\,l\,\sigma}(\textbf{R}_c)
            \cdot\\
        &&  \frac{1}{a^3}\,\int_{a^3} d\textbf{r} \,
            u^*_{\lambda^\prime \textbf{k}^\prime}(\textbf{r})
            [-i\,\hbar\,\nabla] u_{\lambda \textbf{k}}(\textbf{r})\,,
\end{eqnarray*}
where $N$ is the total number of unit cells in the crystal. The
integrand is split thanks to the facts that, in the length of the
unit cell $a$, $\textbf{A}_{{\bf q}\,l\,\sigma}(\textbf{r})$ and
$\exp{[-i (\textbf{k}^\prime-\textbf{k})\cdot \textbf{r}]}$ are
almost constant and $u(\textbf{r})$ is periodic. 
The integral is the momentum matrix element \mbox{${\bf p}_{\lambda^\prime
\textbf{k}^\prime \lambda \textbf{k}}$}. The sum is handled using
the Jacobi-Anger identity, which allows us to separate the angular
from the radial dependences in 
$e^{-i (\textbf{k}^\prime-\textbf{k})\cdot \textbf{R}_c}$, 
and then applying the orthogonality relation for Bessel
functions~\cite{arf}. 
The final expression for the interaction Hamiltonian is
%
\begin{eqnarray}
\label{Eq_H}
H_I &=& -(-i)\,^l
        \,\frac{Q\,A_0}{m}
        \,\frac{1}{L}
        \, e^{-i\,\omega\, t}
        \sum_{\textbf{k} {\textbf{k}}^\prime} \,
        \,\frac{\delta_{\kappa_{\parallel} q_\parallel}}{q_\parallel}
        \,\delta_{\kappa_z q_z}\, \nonumber\\
        &&
        e^{i\,\theta \,l}
        \,\left(\boldsymbol{\epsilon}_\sigma \cdot
        {\bf p}_{c \textbf{k}^\prime v \textbf{k}}\right)
        \, a^\dagger_{c \textbf{k}^\prime} a_{v \textbf{k}} + h.c.  \,,
\end{eqnarray}
%
with \mbox{$\boldsymbol{\kappa} = \textbf{k}^\prime-\textbf{k}$}
having cylindrical coordinates $\{\kappa_\parallel, \theta, \kappa_z
\}$.
\begin{figure}[h]
  \centerline{\includegraphics[scale=0.75]{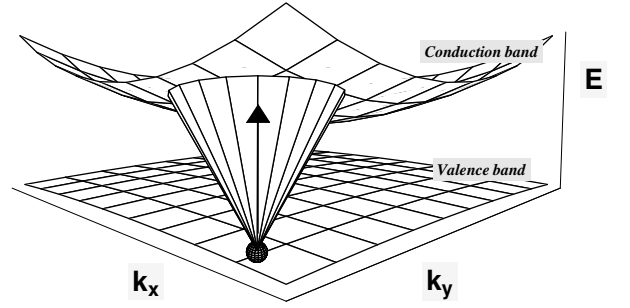}}
  \caption{Schematic representation of interband optical excitation
   with twisted light. A twisted-light beam excites a valence-band
   electron into a superposition of states of the conduction band
   in a ``cone''-like fashion. The laser parameter $q_\parallel$ equals
   the cone radius at the conduction band.
  \label{Fig_Cone}}
\end{figure}
%
%


In order to understand the effect that a twisted-light beam has on
the ground state of a semiconductor, we focus our attention on the
positive part $H_I^{(+)}$ of the interaction Hamiltonian -- first
term of Eq.\ (\ref{Eq_H}). The action of $H_I^{(+)}$ on the full
ground state of the \mbox{$N$-electron} semiconductor can be shown
to yield an eigenvector with expectation value of the orbital
angular momentum equal to $\hbar\,l$. Here we present a simplified
version of the theory where we show the action of the Hamiltonian on
a single-particle valence-band electron state. The analysis is
restricted to optical excitations that couple electron wave-numbers
near $\textbf{k} = 0$, a typical situation in semiconductor optics.
We define \mbox{$\xi = -(-i)\,^l\,(Q\,A_0/m) \,
\left(\boldsymbol{\epsilon}_\sigma \cdot {\bf p}_{c 0 v 0}\right)$},
transform variables $\{ \textbf{k}, \textbf{k}^\prime \} \rightarrow
\{ \textbf{k}, \boldsymbol{\kappa} \}$, fix the value of
$\textbf{k}$ and take the continuum limit for $\boldsymbol{\kappa}$.
In the coordinate representation and rotating frame, the action of
the Hamiltonian on the valence band single-particle initial state
gives
\begin{eqnarray*}
    H_I^{(+)} \varphi_{v \textbf{k}}(\textbf{r})
&=& \,\frac{\xi}{2\,\pi}
    \, \int \,  d\, \boldsymbol{\kappa}  \\
&~~&\frac{\delta(\kappa_{\parallel} - q_\parallel)}{q_\parallel}
    \,\delta(\kappa_z-q_z)
    \,e^{i\,\theta \,l} \, \varphi_{c \,
    \textbf{k}+\boldsymbol{\kappa}}(\textbf{r})
    \,,
\end{eqnarray*}
which makes evident that a ``cone''-like transition occurs producing
a linear superposition of conduction-band states with varying
phases, as depicted in Fig.~\ref{Fig_Cone}. Exploiting the delta
functions, using \mbox{$ \varphi_{\lambda\,
\textbf{k}+\boldsymbol{\kappa}} (\textbf{r}) \simeq
e^{i\,\boldsymbol{\kappa} \,\cdot \,\textbf{r}} \varphi_{\lambda\,
\textbf{k}}(\textbf{r})$}, and applying the Jacobi-Anger identity
\begin{eqnarray}\label{Eq_f}
    H_I^{(+)} \varphi_{v \textbf{k}}(\textbf{r})
&=& \xi \, \left[i^l\,e^{i\,q_z\,z} J_l(q_\parallel
    r_{\parallel})
    \, e^{i\,\phi\,l}\right] \, \varphi_{c \, \textbf{k}}(\textbf{r})
    \nonumber \\
&\doteq&  \xi \,f(\textbf{r}) \, \varphi_{c \, \textbf{k}}(\textbf{r}).
\end{eqnarray}
The dimensionless function $f(\textbf{r}) = i^l\,e^{i\,q_z\,z}
J_l(q_\parallel r_{\parallel}) \, e^{i\,\phi\,l}$ contains all the
relevant information that distinguishes the action of a twisted beam
from its counterpart, the plane-wave. 
Because of the presence of $f(\textbf{r})$, the RHS of Eq.\ (\ref{Eq_f}) 
is {\em not} a conduction-band eigenstate.
For the expectation value of the z-component of the 
orbital angular momentum operator 
$L_z = -i \hbar \partial_{\phi}$ 
in the state of Eq.\ (\ref{Eq_f}) we obtain $<L_z> = \hbar\,l$,
since 
$<L_z> = \hbar\,l$ for the state $e^{i\,\phi\,l}$ 
and $<L_z> = 0$ for the states 
$u_{c \textbf{k}}(\textbf{r})$ and 
$e^{i\textbf{k} \cdot \textbf{r}}$.
Thus, for short times, the evolution leads to a well-defined 
transfer of orbital angular momentum $\hbar\,l$ from the light 
beam to the electron.

This net transfer of orbital angular momentum to the photo-excited
electrons is expected to result in electric currents and associated 
magnetic fields.
The latter would be an experimentally detectable signature of the
type of optical excitation described here, and may also lead
to opto-electronic applications. 
An estimate of the total current/magnetic field produced by all
electrons can be obtained by 
calculating the total number of electrons excited by the field, 
then determining the current/magnetic field of a single electron, 
and, finally, multiplying the single-electron current/magnetic field 
by the number of photo-excited electrons. 
We note that this is a single-particle calculation which does 
not take into account the electron-electron Coulomb
interaction. 
This is justified in the present study by the fact that
the main correction introduced by the Coulomb interaction would be
the renormalization of the single-particle
energies~\cite{hau-jau,fet-wal}. 
This mean-field effect would be
small since we are working in the regime of low excitation (the
density of photogenerated electrons and holes is small), and,
furthermore, by its nature, it does not affect the qualitative
picture drawn in this work. 
Beyond the mean-field approximation, the
Coulomb interaction introduces complex and possibly interesting
scattering effects~\cite{vu-ban-tam-hau, hau-jau}, which are beyond
the scope of our present analysis, and are left for future study.
Again, however, these effects are expected to be small in the
low-excitation regime.
\begin{figure}[h]
  \centerline{\includegraphics[scale=1]{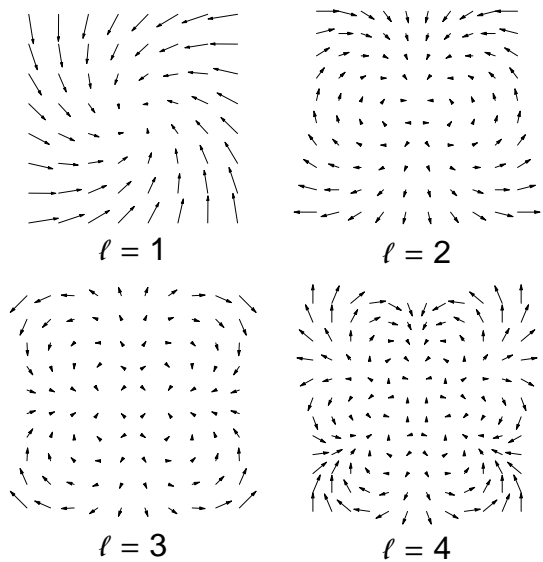}}
  \caption{First-order electric current for OAM $l=1$ to 4.
  The center of each plot coincides with the center of the twisted-light 
  laser beam.
  \label{fig_j1storder}}
\end{figure}

The number of electrons excited in a volume $dV$ and a time interval $t$
is $dN_{ex} = \alpha\, I({\bf r})\, t / \hbar\,\omega$,
with $\alpha$ the absorption coefficient.
$I({\bf r}) = I_0\,e^{-\alpha\,z}$ is the intensity of the light field,
which is related to the vector potential by
\mbox{$ 2\,I_0 = v\,\epsilon \,\omega^{2}\, A_0^{~2}$},
where $v$ is the speed of light in the medium.
Then, the total number of electrons that are excited in $V = L^3$ is
\begin{eqnarray}
    \nonumber
    N_{ex}
&=& \pi\,\frac{L^2 \, t\, I_0}{\hbar\,\omega}\,
    f(q_\parallel\,L)
    \,,
\end{eqnarray}
with $f(q_\parallel\,L)$ a function containing $J_{l-1}(x)$,
$J_{l}(x)$, $J_{l+1}(x)$, and $e^{-\alpha\,L}$.

In order to calculate the electric current, we first determine the
single photo-excited electron state. 
For short times such that
$t(\omega+E_g/\hbar)<1$ the evolution of the state under the action
of the laser field is, in the Schr\"{o}dinger picture,
\begin{eqnarray}\label{Eq_EvolvedState}
    \psi(\textbf{r},t)
&=& U(t)\,\varphi_{v\textbf{k}}(\textbf{r}) \nonumber\\
&\simeq& \varphi_{v\textbf{k}}(\textbf{r}) - e^{-i\,E_g\,t/\hbar}\, (i
    \,t/\hbar) \,\xi\,f(\textbf{r})\,
    \varphi_{c \textbf{k}}(\textbf{r})\,,
\end{eqnarray}
with $U(t)$ the evolution operator (See Eq.~(\ref{Eq_f})). 
The electric-current density of this state is calculated using the
quantum mechanical expression \mbox{${\bf{j}} = (Q\,\hbar/m) \,\Im\,
[\psi^*{\boldsymbol{\nabla}} \psi] - \frac{Q^2}{m} \,\Re\,\left[
\psi^* {\bf A}  \psi\right]$}. 
After a lengthy calculation we arrive
at the electric current and separate the terms in powers of $A_0$
(hidden in $\xi$). 
The zero-order term does not give new phenomena,
and so it is not presented. 
The first-order and second-order terms are
\begin{eqnarray}
    \label{Eq_Current}
    {\bf{j}}_{1}({\bf r},t)
&=& \frac{2\,Q}{m\,L^3}
    \Im\,\left\{\left(\frac{t\,\xi}{\hbar}\right)\,\textbf{p}_{v0c0}
    e^{-i\,E_g\,t/\hbar}\,\, f(\textbf{r}) \right\}\\
    {\bf{j}}_{2}({\bf r},t)
&=& \left(\frac{t\,|\xi|}{\hbar}\right)^2
    \frac{Q\,\hbar}{m\,L^{3}}\,\Im\left\{f(\textbf{r})^*\nabla\,
    f(\textbf{r})\right\}
    \label{Eq_Current_2}
\end{eqnarray}
Let us comment on the main characteristics of these two currents. 
${\bf{j}}_{1}({\bf r},t)$ contains the vector momentum matrix element 
$\textbf{p}_{v0c0}$,
which reappears in the calculation of the current due to the mixing
of the two terms of $\psi(\textbf{r},t)$.
Thus, as $\textbf{p}_{v0c0}$ itself, this component of the macroscopic 
current has a microscopic origin.
This current is somewhat analogous to the optical polarization in 
the standard interband optical transitions induced by plane waves.
Note that this current has no equivalent in atomic physics 
\cite{car-bab-ala-and}.
In Fig~\ref{fig_j1storder} we show the current vector field for
the cases $l=1$ to 4.
A clear feature of the current for $l = 1$ is that it displays
a net circulation around the z-axis, which will give rise to
a sizable magnetic field.
For all other values of $l$ the current shows net circulations 
around off-centered axes, like in the cases $l=2$ to 4 seen 
in Fig~\ref{fig_j1storder}, but no net circulation around the
center of the beam.
Finally, without explicitly showing it here, we mention that the
current field behaves like a travelling wave in the z-direction, 
and therefore, at a given time, clockwise and counterclockwise 
circulations alternate as one moves along the z-axis. 
The second-order term ${\bf{j}}_{2}({\bf r},t)$ comes from just 
the second term in $\psi(\textbf{r},t)$. 
As can be seen in Eq.~(\ref{Eq_Current_2}), ${\bf{j}}_{2}({\bf r},t)$
stems directly from the function $f(\textbf{r})$, defined after
Eq.~(\ref{Eq_f}), whose spatial dependence mimics closely that of laser 
beam and varies significantly only on a macroscopic scale.
Thus, we interpret that this current has a macroscopic origin and is 
analog to the motion induced by twisted light on atomic systems.

For the currents with net circulation around the beam axis, 
e.g.\ the case $l=1$ in Fig.\ \ref{fig_j1storder}, we estimate 
the magnetic field $b_z$ at the center the beam.
We adopt a semiclassical approach and use Biot-Savart's law with 
the currents from Eq.~(\ref{Eq_Current}) and (\ref{Eq_Current_2}) 
whose azimuthal components are
\begin{eqnarray}\label{Eq_j1approx}
    {j}_{1 \phi}({\bf r},t)
&=& -2\,\frac{q^2\,A_0\,p_0^2}{m^2\,L^3\,\hbar}
    \, J_1(q_\parallel\,r_\parallel)\,\sin\left(q_z\,z\right)\, t \, ,
    \\
    {j}_{2 \phi}({\bf r},t)
&=& l\,\left(\frac{t\,|\xi|}{\hbar}\right)^2 \,
    \frac{q\,\hbar}{m\,L^{3}}\,
    \frac{1}{r_\parallel}\,J_l(q_\parallel\,r_\parallel)^2 \, .
\end{eqnarray}
Then, the total magnetic field is determined by $B_z = N_{ex} \, b_z$.
Our semiclassical approach is an admissible procedure in perturbation 
theory since the quantum fluctuations of the current are of higher order 
in the vector potential and thus smaller than their mean values.

As an example, we estimate the magnetic field $B_z$ for a realistic 
situation. 
We choose the following typical experimental parameters
\cite{sim-nyg-hu,ala-bab,tab-pet,one-mac-all-pad}. 
For the laser beam: Ti:Saphire laser with frequency 
\mbox{$\lambda = 589$ nm}, 
repetition rate \mbox{$F = 100$ MHz}, 
power \mbox{$P = 1\, \mu$J s$^{-1}$}, 
pulse duration \mbox{$t = 10$ fs},
\mbox{$q_{z} = 10^{7}$ m$^{-1}$},
\mbox{$q_\parallel = 4\,10^{6}$ m$^{-1}$} 
(spot size \mbox{$\simeq(0.5 \,\mu$m$)^2$}), 
and $l=1$. 
For the semiconductor material we choose GaAs parameters: 
\mbox{$\alpha = 10^5$ m$^{-1}$}, 
\mbox{$a = 0.6$ nm},
\mbox{$v = c/3$}, 
\mbox{$L = 1.3\,\mu$m} and 
\mbox{$|\boldsymbol {\epsilon}_\sigma \cdot {\bf p}_{c 0 v 0}| = 
6.4\,10^{-25}$ kg m/s}.
Assuming a top-hat laser pulse (which approximates the cw-field of
Eq.~(\ref{Eq_H}) for long enough pulses) these values yield 
\mbox{$N_{ex} = 2\, 10^{5}$} and
\mbox{$(|\xi| \, t / \hbar) = 0.95$}.
The resulting z-component of the magnetic field in the center of the 
beam is 
\mbox{$|B_{1z}| = 1.2 \, $mT} and 
\mbox{$|B_{2z}| = 0.4 \, \mu$T} 
for the first-order and second-order currents, respectively.
This estimate is reliable since the build-up of the magnetic field
takes place in a time shorter than the typical relaxation times of
electrons in the conduction band of semiconductors at low excitation
density~\cite{hau-jau}.

Our theory can be put to test by measuring this predicted magnetic
field.  
We suggest the use of time-resolved Faraday rotation.
A quick estimate of the polarization rotation angle can be obtained
with the Verdet formula $\theta = V B_z \ell$, where $V$ is the Verdet
constant and $\ell$ is the length traversed by the probe beam.
Taking $V= 10^{-4} \, \text{rad/(G\, cm)}$ and $\ell = 10 \, \mu$m 
one obtains $\theta \simeq 10^{-6}\, \text{rad}$, which is clearly measurable 
according to Ref.\ \cite{kik-smo-sam-aws}.


In conclusion, in this Letter we have shown for the first time the effect
that light carrying orbital angular momentum has on semiconductors.
We started by generalizing the basic theory of photo-excitation
with plane waves to the case of Bessel mode beams.
We found that the excitation process generates a superposition of
conduction-band states with a well-defined angular momentum.
This superposition state is associated to an electric-current 
density, which acts as the source of a magnetic field.
As an example, the magnitude of this magnetic field has been estimated
in a typical experimental situation.
The important regime of excitonic generation has not been addressed here
and will be the subject of a future work, as well as the inclusion of
electron-electron Coulomb interaction effects.

We acknowledge financial support from ANPCyT (PICT-11609/3),
University of Buenos Aires (UBACyT X179), and CONICET (PIP-5851).
P.I.T.\ is a researcher of CONICET.


\begin{thebibliography}{99}

\bibitem{bet}
R.\ A.\ Beth,
Phys.\ Rev.\ {\bf 50}, 115 (1936).

\bibitem{all-bei-spr}
L.\ Allen, M.\ W.\ Beijersbergen, R.\ J.\ C.\ Spreeuw, and J.\ P.\ Woerdman,
Phys.\ Rev. A {\bf 45}, 8185 (1992).

\bibitem{mol-tor-tor}
G.\ Molina-Terriza, J.\ P.\ Torres, and L.\ Torner,
Nature Phys.\ {\bf 3}, 305 (2007).

\bibitem{pad-cou-all}
M.\ Padgett, J.\ Courtial, and L.\ Allen,
Phys.\ Today {\bf 57}, Iss.\ 5, 35 (2004).

\bibitem{all-pad-bab}
L.\ Allen, M.\ J.\ Padgett, and M.\ Babiker,
Prog.\ Opt.\ {\bf XXXIX}, 291 (1999).

\bibitem{bar-tab}
S.\ Barreiro and J.\ W.\ R.\ Tabosa,
Phys.\ Rev.\ Lett.\  {\bf 90}, 133001 (2003).

\bibitem{all}
B.\ Allen,
J.\ Opt.\ B.\ Quantum Semiclass.\ Opt.\ \textbf{4}, S1-S6 (2002).

\bibitem{fri-nie-hek-rub}
M.\ E.\ J.\ Friese, T.\ A.\ Nieminen, N.\ R.\ Heckenberg, and H.\
Rubinsztein-Dunlop,
Nature \textbf{394}, 348 (1998).

\bibitem{mut-str}
B.\ A.\ Muthukrishnan and C.\ R.\ Stroud Jr.,
J.\ Opt.\ B.\ Quantum Semiclass.\ Opt.\ \textbf{4}, S73-S77 (2002).

\bibitem{dav-and-bab}
B.\ L.\ C.\ D\'avila-Romero, D.\ L.\ Andrews, and M.\ Babiker,
J.\ Opt.\ B.\ Quantum Semiclass.\ Opt.\ {\bf 4}, S66-S72 (2002).

\bibitem{ara-ver-cla}
F.\ Araoka, T.\ Verbiest, K.\ Clays, and A.\ Persoons,
Phys.\ Rev.\ A \textbf{71}, 055401 (2005).

\bibitem{ala-bab}
S.\ Al-Awfi and M.\ Babiker,
Phys.\ Rev.\ A {\bf 61}, 033401 (2000).

\bibitem{and-ryu-cla}
M.\ F.\ Andersen, C.\ Ryu, P.\ Clade, V.\ Natarajan, A.\ Vaziri,
K.\ Helmerson, and W.\ D.\ Phillips,
Phys.\ Rev.\ Lett.\  \textbf{97}, 170406 (2006).

\bibitem{sim-nyg-hu}
T.\ P.\ Simula, N.\ Nygaard, S.\ X.\ Hu, L.\ A.\ Collins,
B.\ I.\ Schneider, and K.\ Molmer,
arXiv:0707.3698v1 [cond-mat.soft].

\bibitem{jau}
R.\ J\'{a}uregui,
Phys.\ Rev.\ A \textbf{70}, 033415 (2004).

\bibitem{hau-koc}
H.\ Haug and S.\ W.\ Jauho,
{\it Quantum theory of the optical and electronic properties of semiconductors,
Fourth Edition} (World Scientific Publishing Company, Singapore, 2004).

\bibitem{arf}
G.\ Arfken,
{\it Mathematical methods for physicists, Third Edition}
(Academic Press, Inc., Orlando, 1985).

\bibitem{hau-jau}
H.\ Haug and A.-P.\ Jauho,
{\it Quantum kinetics in transport and optics of semiconductors, Second Edition}
(Springer-Verlag, Berlin, Heidelberg, 2007).

\bibitem{fet-wal}
A.\ L.\ Fetter and J.\ D.\ Walecka,
{\it Quantum Theory of Many-Particle Systems}
(Dover Publications, 2003).

\bibitem{vu-ban-tam-hau}
Q.\ T.\ Vu, L.\ Banyai, P.\ I.\ Tamborenea, and H.\ Haug,
Europhys. Lett. 40, 323 (1997).

\bibitem{car-bab-ala-and}
A.\ R.\ Carter, M.\ Babiker, M.\ Al-Amri and D.\ L.\ Andrews,
Phys.\ Rev.\ A {\bf 73}, 021401 (2006).

\bibitem{tab-pet}
J.\ W.\ R.\ Tabosa and D.\ V.\ Petrov,
Phys.\ Rev.\ Lett.\ {\bf 83}, 004967 (1999).

\bibitem{one-mac-all-pad}
A.\ T.\ O'Neil, I.\ MacVicar, L.\ Allen, and M.\ J.\ Padgett,
Phys.\ Rev.\ Lett.\ {\bf 88}, 053601 (2002).

\bibitem{kik-smo-sam-aws}
J.\ M.\ Kikkawa, I.\ P.\ Smorchkova, N.\ Samarth, and D.\ D.\ Awschalom,
Science {\bf 277}, 1284 (1997).

\end{thebibliography}
\end{document}